\documentclass[twocolumn]{emulateapj}
\begin{document}

\title{A LACK OF RESOLVED NEAR-INFRARED POLARIZATION ACROSS THE FACE OF M51}
\author{Michael D. Pavel\altaffilmark{1} and Dan P. Clemens}
\altaffiltext{1}{Current address: Department of Astronomy, The University of Texas at Austin, 1 University Station, C1400, Austin, TX 78712-0259}
\affil{Institute for Astrophysical Research}
\affil{Boston University, 725 Commonwealth Ave, Boston, MA 02215}
\email{pavelmi@astro.as.utexas.edu; clemens@bu.edu}

\slugcomment{Accepted for Publication in The Astrophysical Journal Letters}

\shorttitle{M51 Polarimetry}
\shortauthors{Pavel, M.~D. \& Clemens, D.~P.}

\begin{abstract}

The galaxy M51 was observed using the Mimir instrument on the Perkins telescope to constrain the resolved H-band (1.6 $\mu$m) polarization across the galaxy. These observations place an upper limit of $P_H<0.05\%$ on the $H$-band polarization across the face of M51, at 0.6 arcsecond pixel sampling. Even with smoothing to coarser angular resolutions, to reduce polarization uncertainty, the $H$-band polarization remains undetected. The polarization upper limit at $H$-band, when combined with previous resolved optical polarimetry, rules out a Serkowski-like polarization dependence on wavelength. Other polarization mechanisms cannot account for the observed polarization ratio ($P_H/P{VRI} \lesssim 0.05$) across the face of M51.

\end{abstract}

\keywords{Galaxies: individual: (M51) --- Galaxies: ISM --- Galaxies: magnetic fields --- Infrared: galaxies --- Polarization --- Radiation mechanisms: general}

\section{Introduction}

A comparative study of magnetic fields probed at optical, near-infrared (NIR), and radio wavelengths for a face-on galaxy can elucidate the nature of galaxy-scale magnetic fields and provide context for studies of magnetic fields in the Milky Way. To this end, the galaxy M51 was chosen for a pilot study of resolved polarimetry with the Mimir instrument \citep{2007PASP..119.1385C}. M51 is an essentially face-on, grand-design spiral galaxy, though not a perfect Milky Way analog because it is interacting with NGC 5195 \citep{1972ApJ...178..623T}. Its high elevation (declination $+47\degr$) made it suitable for multiple, long observations and its large angular size ($\sim 11 \times 7$ arcmin) was well matched to the Mimir $10 \times 10$ arcmin field-of-view. 

How does the magnetic pitch angle across the face of M51 vary with position at NIR wavelengths, and how does this compare to pitch angle variations seen at other wavelengths \citep{2006A&A...458..441P}? How do these pitch angles compare to results from the Milky Way \citep[e.g.,][]{1994A&A...288..759H,1996ApJ...462..316H,2011ApJ...738..192P,2012ApJ...749...71P}? Since optical/NIR polarization traces magnetic fields in dusty regions while synchrotron polarization and Faraday rotation trace magnetic fields in the warm interstellar medium, are the same large-scale magnetic fields being observed in both regions, or are local magnetic fields able to decouple from the large-scale field under certain circumstances? Does the resolved polarization across the face of M51 follow a Serkowski dependence of polarization strength with wavelength \citep{1975ApJ...196..261S} as seen for Galactic lines-of-sight?

The magnetic field of M51 has been previously studied at optical and radio wavelengths, but never at NIR wavelengths. The first observations of the magnetic field in M51 were made by \citet{1972A&A....17..468M} at 1415 MHz ($\approx$ 21 cm) using synchrotron emission. Later, \cite{1976Natur.264..222S} observed polarized synchrotron emission at 6 and 21 cm, but found no measurable polarization at 49 cm. The observed polarizations suggested that the large-scale magnetic field is oriented parallel to the spiral arms of M51. Additional polarimetric observations by \citet{1992A&A...263...30N} at 2.8 cm, \citet{1992A&A...265..417H} at 18 and 20.5 cm, \citet{2009A&A...503..409H} at 18 and 22 cm, and \citet{2011MNRAS.412.2396F} at 3 and 6 cm support this interpretation. However, in all cases, those authors warn that Faraday depolarization may affect interpretation of the observations.

To probe the magnetic field of M51 without the effects of Faraday depolarization, \citet{1987MNRAS.224..299S} used resolved optical polarimetry ($P_{VRI}$: 450-1000 nm, with peak response at 850 nm) to measure the orientation of disk magnetic fields (as projected onto the plane of the sky) assuming the polarization was caused by magnetically-aligned dust grains. Those observations also traced an open spiral pattern from the nucleus to a radial offset from the core of 4-5 kpc.

This pilot study intended to extend the work of \citet{1987MNRAS.224..299S} by measuring resolved polarimetry across the disk of M51 in the NIR. However, the fully reduced H-band polarimetric observations of M51 showed \textit{a lack} of polarization. After coadding multiple observations, a very low upper limit on the H-band polarization was set. A number of polarization mechanisms are considered, but none can account for the observed and absent polarizations.

\begin{figure*}[ht]
	\centering
		\includegraphics[scale=.8]{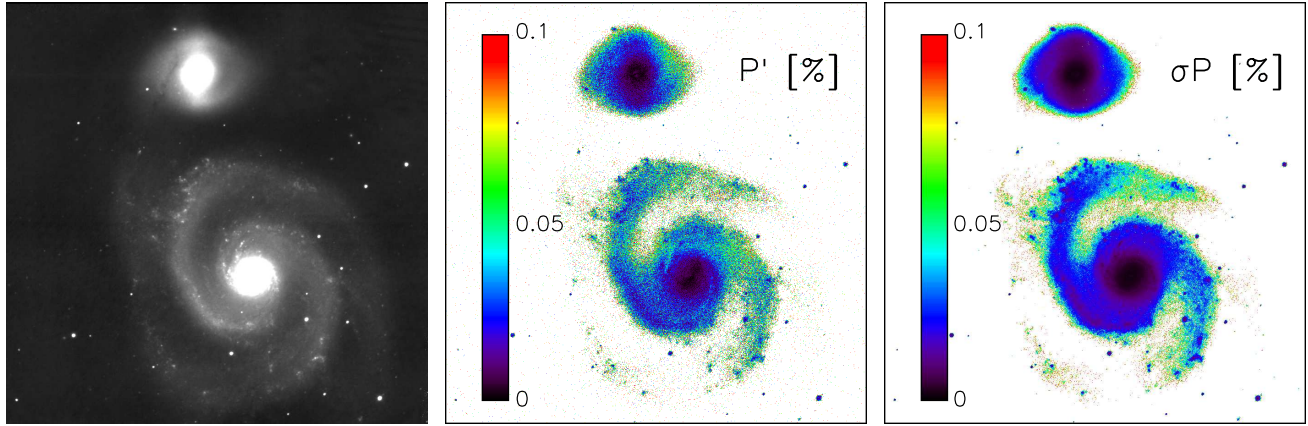}
	\caption[H-band Polarimetry of M51]{\label{M51_image}(Left Panel) Linear display of the Mimir $9\times9$ arcmin H-band photometric image of M51. This image is the result of the coadding of all observations of M51. (Middle panel) Measured polarization before statistical debiasing (P$'$) at the native 0.6 pixel sampling. (Right panel) Measured polarization uncertainty across M51 at the same sampling.}
\end{figure*}

\section{Observations and Analysis}

M51 was observed in 2009 February, March, and May with Mimir on the Perkins 1.8m Telescope outside of Flagstaff, AZ. Fifty observations were obtained, totaling 14.5 hours of on-target integration. Imaging polarimetry was measured by a combination of a cold, stepping half-wave plate (HWP, to modulate the polarization signal) and a cold, fixed wire grid. For each pointing, the HWP was rotated to 16 unique position angles (the first HWP position was measured twice per HWP rotation, for a total of 17 images). At each HWP angle, a 10.25 sec exposure was taken. This procedure was repeated toward six different sky dither positions, offset by 15 arcsec, following a rotated hexagon pattern on the sky, for a total of 102 images per pointing. Resolved polarimetry with Mimir in this manner has successfully been used for the comet 73P/Schwassmann-Wachmann 3 \citep{2008AJ....135.1318J} and for the reflection nebula IRAS 05329-0505 \citep{dissertation}. The Mimir data reduction packages (developed for starlight polarimetry) described in \citet{GPIPS_II,GPIPS_I} were applied to these observations, but with the key differences described below.

\subsection{Background Fitting}

To remove sky backgrounds, the Mimir reduction software normally calculates `super-sky' images for each HWP position after flat-fielding, to remove the second-order (illumination and other) effects. To do so, for each HWP position during a single observation, all images taken through that HWP position are coadded without astrometric registration to reject stars and the result is subtracted from the constituent HWP images. Since the sky background may be polarized, it is important to preserve this polarization information by calculating such sky backgrounds independently for each HWP angle. For M51, super-sky images created in this manner would end up removing much of the object's flux (since the 15 arcsec dither size is smaller than M51). Instead, backgrounds were modeled and removed using a different method.

For each pointing, `Sky only' zones were identified by eye in the Mimir images of M51 and used to fit two-dimensional, second-order polynomials to the sky background across each image. While these zones were identical in all 102 images per observation, the fitted sky background polynomials were independently fit in each image. These zones avoided M51, NGC 5195, and the brightest stars in the field. The choice of a second-order polynomial was based on an analysis of the sky background residuals. These polynomial fits were then subtracted from the images to remove the sky background prior to astrometric registration and co-adding.

\subsection{Resolved Polarimetry}

The analysis of resolved polarimetry is similar to the analysis described in \citet{GPIPS_I} for unresolved starlight polarimetry. After removing the sky background, images were grouped by HWP angle and coadded to form 16 master HWP images for each observation. Since the observations were dithered, this step removed image defects (e.g., bad pixels and cosmic rays). Because of the 4$\theta$ modulation of the polarization as the HWP rotated, the 16 master HWP angles correspond to four independent measurements of each of the instrument independent polarimetric position angles (IPPAs): 0, 45, 90, and 135$\degr$. The four images at each IPPA were averaged to form four master IPPA images per observation.

From these four master IPPA images, corrected Stokes images ($Q/U_{\rm CORR}$) were calculated:
\begin{equation}
	\label{q_equ}
	Q_{\rm CORR} = (I_0 - I_{90})/(I_0 + I_{90}) - Q_{\rm INST}
\end{equation}
\begin{equation}
	\label{u_equ}
	U_{\rm CORR} = (I_{45} - I_{135})/(I_{45} + I_{135}) - U_{\rm INST},
\end{equation}
where operations were performed on each pixel in the entire image and $Q/U_{\rm INST}$ are the previously determined instrumental polarization contributions \citep{GPIPS_II}.

These corrected images were scaled by the polarization efficiency ($\eta$ = $91.1\pm0.4\%$) and rotated by the HWP zero-phase offset angle \citep[$\theta$, the angle between the instrument coordinate system and equatorial;][]{GPIPS_II}:
\begin{equation}
	\label{q_rot}
	Q = cos(2\theta)Q_{\rm CORR}/\eta - sin(2\theta)U_{\rm CORR}/\eta
\end{equation}
\begin{equation}
	\label{u_rot}
	U = cos(2\theta)U_{\rm CORR}/\eta + sin(2\theta)U_{\rm CORR}/\eta.
\end{equation}

This generated a total of 50 unique Stokes $U$ and 50 Stokes $Q$ images for M51. These were astrometrically registered and coadded to produce the final Stokes $U$ and $Q$ images for M51. From these corrected, calibrated Stokes images, an image of the biased degree of polarization ($P'$) was calculated:
\begin{equation}
	\label{p_equ}
	P' = \sqrt{U^2 + Q^2}
\end{equation}
Since the degree of polarization suffers from a positive statistical bias, $P'$ was debiased using the prescription of \citet{1974ApJ...194..249W}, to calculate the true polarization:
\begin{equation}
	P = \sqrt{P'^2 -\sigma_P^2}.
\end{equation}

A deep photometric image was also calculated by registering and coadding all of the individual observations and is shown in the left panel of Figure \ref{M51_image}.

\section{Results}

The middle panel of Figure \ref{M51_image} shows the biased polarization percentage at Mimir's native 0.6 arcsecond pixel sampling before statistical debiasing and before any attempt was made to increase polarization signal-to-noise ratio (SNR) via smoothing. After applying the correction for the positive statistical bias, the polarization uncertainties (Fig. \ref{M51_image}; right panel) are greater than the measured polarizations across the entire image. Therefore, there is \textit{no measurable H-band polarization across M51 or NGC 5195}. A polarization position angle image was also calculated, but is meaningless when there is no measurable polarization and so is not shown.

The uncertainty in the measured degree of polarization, at 0.6 arcsecond sampling, is shown in the right panel of Figure \ref{M51_image}. The polarization uncertainty is almost perfectly anticorrelated with the total flux, shown in the left panel of Fig. \ref{M51_image}. This is expected, since the polarization uncertainty depends directly on the uncertainty in the flux measured at each HWP angle. If photon noise dominates, then the photometric SNR should increase with total flux. Additional sources of uncertainty include the background subtraction, a conservative 0.1\% instrumental polarization calibration uncertainty, and the uncertainty in the polarization efficiency (1 part in 225).

\begin{figure}
	\centering
		\includegraphics[scale=.52]{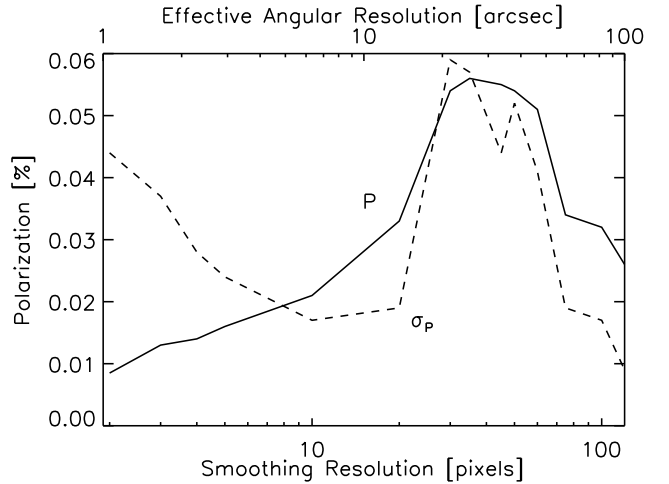}
	\caption[]{\label{smoothing}Runs of median polarization percentage (P), and its uncertainty($\sigma_{\rm P}$), versus smoothing resolution, for all M51 pixels showing emission brighter than the sky background. Median polarization is shown as the solid curve; uncertainty as the dashed curve. The lower x-axis represents the number of pixels (n) combined into a new (n$\times$n) synthetic pixel. The upper x-axis is the corresponding angular resolution of the new synthetic pixel. Stronger smoothing increases the median polarization, but also increases its uncertainty, thereby failing to improve the ratio of those quantities.}
\end{figure}

Since the Stokes $U$ and $Q$ parameters are Gaussian-distributed quantities, the polarimetric SNR across the face of M51 can be increased by smoothing Stokes $U$ and $Q$ images to synthesize lower angular resolution images. The effects of smoothing on debiased polarization (P) and its uncertainty ($\sigma_P$) across the face of M51 are shown in Figure \ref{smoothing} for the median of pixels with emission above the mean sky background. Smoothing to 5.8 arcsecond resolution (10 pixels) caused the polarization uncertainty to decrease and the debiased polarization to increase. Despite this, the median debiased degree of polarization remained consistent with no detection. Smoothing to even coarser resolutions (15-60 arcsec) resulted in larger polarizations, but also larger corresponding uncertainties. Smoothing beyond 60 arcsec caused both the polarization and uncertainty to fall, but the SNR never exceeded two.

Smoothing to such large resolutions will tend to mask any bona fide polarization signal because of depolarization. If the spiral-like pattern seen by \citet{1987MNRAS.224..299S} exists at H-band, these last large smoothing kernels would have smoothed over regions with significant rotation in PA and would effectively depolarize the synthetic pixels. This is critical in the core of M51, where a large synthetic pixel would encompass all possible PAs. Therefore, the marginal polarization detections at resolutions larger than 60 arcseconds are likely not reliably revealing polarizations across the face of M51.

\section{Discussion}

These results place an H-band polarization $1\sigma$ upper limit of 0.05\% across the face of M51, at an angular sampling of 0.6 arcseconds and 0.02\% at a resolution of 1.8 arcseconds. Low NIR polarizations are expected, given the $\sim 1\%$ optical polarizations from \citet{1987MNRAS.224..299S}.  If the polarization is caused by dichroic dust extinction in the disk of M51, then the wavelength dependence of polarization should follow a Serkowski law \citep{1975ApJ...196..261S,1982AJ.....87..695W}:
\begin{equation}
\frac{P_{\lambda}}{P_{\rm max}} = exp \left[ -(1.86\,\lambda_{\rm max} - 0.10) ln^2 \left( \frac{\lambda_{\rm max}}{\lambda} \right) \right],
\end{equation}
where $\lambda_{\rm max}$ is the wavelength of maximum polarization. Assuming $\lambda_{\rm max} = 0.55 \,\mu$m \citep[typical for Galactic lines-of-sight;][]{1975ApJ...196..261S} would predict $P_{\rm H}/P_{VRI} = 0.39$, whereas the observations show $P_{\rm }H/P_{VRI} < 0.05\pm 0.05$.

To test whether \textit{any} Serkowski Law could account for these observations, the assumption of $\lambda_{\rm max}= 0.55\,\mu$m was relaxed and the NIR and optical polarization data were used to fit for the best $\lambda_{\rm max}$, using the MPFITFUN routine from the IDL-based MPFIT package \citep{2009ASPC..411..251M}. No value of $\lambda_{\rm max}$ could recreate the observed polarization ratio. Polarization by dust dichroism, which is observed throughout the Milky Way \citep{1970MmRAS..74..139M}, cannot explain these observations.

Other possible polarization sources were considered to try to account for the extra optical polarization, including electron (Thompson) scattering of low energy photons. However that polarization mechanism is wavelength independent \citep{1985ApJ...297..621A}, so any observed wavelength dependence would only reflect the polarization spectrum of the photon source. The significant observed difference between the degree of polarization at optical and NIR wavelengths rules out electron scattering, since it would only enhance the degree of polarization, not change $P_{\rm H}/P_{\rm VRI}$. Polarized synchrotron radiation is also unable to account for the observed polarization ratio, since the degree of polarization only depends on the electron power law distribution. The synchrotron intensity ratio at optical and H-band would have to be at least $1\% / 0.05\%=20$. This requires either optically thick synchrotron emission or optically thin emission with an inverted electron energy distribution (both are unlikely).

For now, the cause of the observed polarization at optical and the lack of polarization at NIR wavelengths across the face of M51 remains unknown. Resolved polarimetry in multiple passbands may provide additional insight, especially at optical wavelengths. However, a current lack of instrumentation with resoved imaging polarimety capabilities limits these possibilities.

\acknowledgements

The authors thank T. J. Jones for useful discussions. This research was conducted in part using the Mimir instrument, jointly developed at Boston University and Lowell Observatory and supported by NASA, NSF, and the W.M. Keck Foundation. The GPIPS effort has been made possible by grants AST 06-07500 and AST 09-07790 from NSF/MPS to Boston University and the Boston University-Lowell Observatory partnership.

{\it Facilities:} \facility{Perkins}

\end{document}